\documentclass[final,3p,times]{elsarticle}

\usepackage{amssymb}
\usepackage{hyperref}
\usepackage{rotating}
\usepackage{xcolor}

\hypersetup{linkcolor=red,citecolor=magenta,filecolor=cyan,urlcolor=magenta}
\usepackage{amsmath}
\usepackage{url}
\usepackage{chapterbib}

\newcommand\arcsec{\mbox{$^{\prime\prime}$}}%

\usepackage{etoolbox} 
\makeatletter

\def\elsReceived{\@empty}
\def\elsRevised{\@empty}
\def\elsAccepted{\@empty}
\newcommand\received[1]{\gdef\elsReceived{#1}}
\newcommand\revised[1]{\gdef\elsRevised{#1}}
\newcommand\accepted[1]{\gdef\elsAccepted{#1}}

\newcommand\elsprintdates{%
  \ifx\elsReceived\@empty\else
    {\small
     \noindent\textit{Received:} \elsReceived
     \ifx\elsRevised\@empty\else; \textit{Revised:} \elsRevised\fi
     \ifx\elsAccepted\@empty\else; \textit{Accepted:} \elsAccepted\fi
     \par\vskip10pt}%
  \fi
}

\patchcmd{\MaketitleBox}%
  {\footnotesize\itshape\elsaddress\par\vskip36pt}%
  {\footnotesize\itshape\elsaddress\par\vskip6pt\elsprintdates\vskip30pt}%
  {}{\typeout{*** elsarticle patch failed ***}}

\makeatother

\journal{Science Bulletin}

\begin{document}

\begin{frontmatter}



\title{A Jetted Wandering Massive Black Hole Candidate in a Dwarf Galaxy}

\author[label1]{Yuanqi Liu}
\author[label1,label2,label3]{Tao An \corref{cor1}}
\author[label4,label5]{Mar Mezcua}
\author[label1]{Yingkang Zhang}
\author[label6]{Ailing Wang}
\author[label7]{Jun Yang}
\author[label8]{Xiaopeng Cheng}

\affiliation[label1]{
  organization={Shanghai Astronomical Observatory},
  organization={Chinese Academy of Sciences},
  city={Shanghai},
  postcode={200030},
  country={China},
}

\affiliation[label2]{
  organization={School of Astronomy and Space Sciences},
  organization={University of Chinese Academy of Sciences},
  city={Beijing},
  postcode={100049},
  country={China},
}

\affiliation[label3]{
  organization={State Key Laboratory of Radio Astronomy and Technology},
  city={Beijing},
  postcode={100101},
  country={China},
}

\affiliation[label4]{
  organization={Institute of Space Sciences (ICE, CSIC)},
  organization={Campus UAB},
  city={Barcelona},
  postcode={08193},
  country={Spain},
}

\affiliation[label5]{
  organization={Institut d'Estudis Espacials de Catalunya (IEEC)},
  organization={Edifici RDIT, Campus UPC},
  city={Castelldefels (Barcelona)},
  postcode={08860},
  country={Spain},
}

\affiliation[label6]{
  organization={Key Laboratory of Particle Astrophysics},
  organization={Institute of High Energy Physics},
  organization={Chinese Academy of Sciences},
  city={Beijing},
  postcode={100049},
  country={China},
}

\affiliation[label7]{
  organization={Department of Space, Earth and Environment},
  organization={Chalmers University of Technology},
  organization={Onsala Space Observatory},
  city={Onsala},
  postcode={SE-43992},
  country={Sweden},
}

\affiliation[label8]{
  organization={Korea Astronomy and Space Science Institute},
  city={Daejeon},
  postcode={34055},
  country={Korea},
}

\cortext[cor1]{Corresponding author. 
  E-mail: antao@shao.ac.cn}

\received{24-Feb-2025}

\revised{09-Jun-2025}

\accepted{21-Aug-2025}
\end{frontmatter}

Dwarf galaxies, with their shallow gravitational potentials, less evolved assembly histories, and low mass, serve as critical laboratories for studying massive black hole (MBH) formation and growth \citep{2020ARA&A..58..257G}. They preserve signatures of the primordial processes that shaped early MBH emergence\citep{2020ARA&A..58...27I}.
Off-nuclear(or offset) active galactic nucleus (AGN) are increasingly recognized as important laboratories for understanding galactic dynamics and black-hole evolution, with recent studies indicating that they are quite common in various galaxy populations. For instance, systematic surveys using optical spectroscopy, X-ray, and integral field unit (IFU) observations have identified spatially or kinematically offset AGNs in 2\%--62\% of samples, depending on methodology \citep{2015ApJ...806..219C, 2016ApJ...829...37B, 2024MNRAS.528.5252M}. Regarding AGNs with offset radio cores, Popkov et al. \citep{2025arXiv250617769P} identified $\sim35$ cases where Very Long Baseline Interferometry (VLBI) coordinates are associated with bright jet components separated by several to tens of milliarcseconds (mas) from the radio core. The primary mechanisms for such off-nuclear AGNs include dual/binary supermassive black holes (SMBHs) and jet-driven displacements, though systems hosted in dwarf galaxies remain rare and more complex due to their shallow potentials and merger histories.
The discovery of off-nuclear MBHs challenges traditional evolutionary models that typically posit nuclear gas reservoirs as the primary sites of MBH growth. MBH displacement from the galactic centre can result from gravitational wave recoil during mergers or asymmetric gas accretion \citep{2004ApJ...607L...9M}. Recoil velocities can even exceed the escape velocities of dwarf galaxies. Numerical simulations indicate that approximately 50\% of MBHs in dwarf galaxies might reside more than 1 kiloparsec (kpc) from their hosts' centre \citep{2021MNRAS.505.5129B},  a prediction consistent with current observations \citep{2024MNRAS.528.5252M, 2020ApJ...888...36R}.

However, detecting these displaced MBHs remains challenging, particularly in low-mass systems where shallower gravitational potentials increase their likelihood of ejection. The Mapping Nearby Galaxies at Apache Point Observatory (MaNGA) survey has advanced AGN detection in dwarf galaxies through spaxel-by-spaxel emission-line analysis, surpassing the traditional single-fiber spectroscopy limitations. Using spatially resolved Baldwin-Phillips-Terlevich (BPT) diagnostics with [N II], [S II], and [O I] line ratios, MaNGA has identified AGN candidates in 628 dwarf galaxies \citep{2024MNRAS.528.5252M, 2025MNRAS.536..295M}. Notably, 62\% of these systems show offsets $\geq 3$ arcseconds ($^{\prime\prime}$) between AGN-dominated spaxels and the galactic centre. At $z\sim$0.03, MaNGA's resolution of about 1 kpc allows differentiation between three scenarios: light echoes from past AGN flares, wandering intermediate-mass black holes (IMBHs) displaced by dynamical interactions, and obscured low-accretion AGNs \citep{2020ApJ...898L..30M}. Recent extended ROentgen Survey with an Imaging Telescope Array (eROSITA) X-ray data \citep{2024ApJ...974...14S} corroborated with MaNGA findings and theoretical predictions, reveals that roughly 50\% of eROSITA-detected AGNs are spatially off-nuclear.

However, strong emission from H II regions in star-forming dwarfs often obscures AGN optical signatures, with [O III]/H$\beta$ ratios mimicking AGN excitation \citep{2023MNRAS.521.1264T}. Ultraluminous X-ray sources (ULXs) can also imitate AGN variability, hindering X-ray diagnostics. Radio observations, especially VLBI, prove  invaluable by detecting compact cores with brightness temperatures ($T_{\rm b} > 10^9$~K), exceeding star-forming maxima ($\sim 10^7$ K) \citep{2020ApJ...888...36R}.

Recent Very Long Baseline Array (VLBA) observations of 13 radio-selected dwarf galaxy candidates detected four compact mas-scale structures in galaxy outskirts with $T_{\rm b}>10^{6}$ K \citep{2022ApJ...933..160S}, consistent with AGN but likely background sources due to large offsets. Remaining sources could have extended radio emission on scales of tens of parsecs, which could originate from supernova remnants or AGN-powered radio lobes \citep{2011Natur.470...66R}.

Previous surveys lacked the astrometric precision needed to unambiguously associate compact radio sources with dwarf galaxies, leading to significant background contamination \citep{2022ApJ...933..160S}. Optical and X‑ray diagnostics also struggle to separate faint AGN from ultraluminous X‑ray binaries in galaxy outskirts \citep{2024A&A...682A..36H}. Finally, only a few off‑nuclear MBHs with sustained accretion have been confirmed, limiting our understanding of early black‑hole growth. Although recent observations have expanded the census of AGN candidates in dwarf galaxies \citep{2025ApJ...982...10P}, VLBI confirmations remain scarce.

\begin{figure}[htbp]
    \centering
    \includegraphics[width=0.9\textwidth]{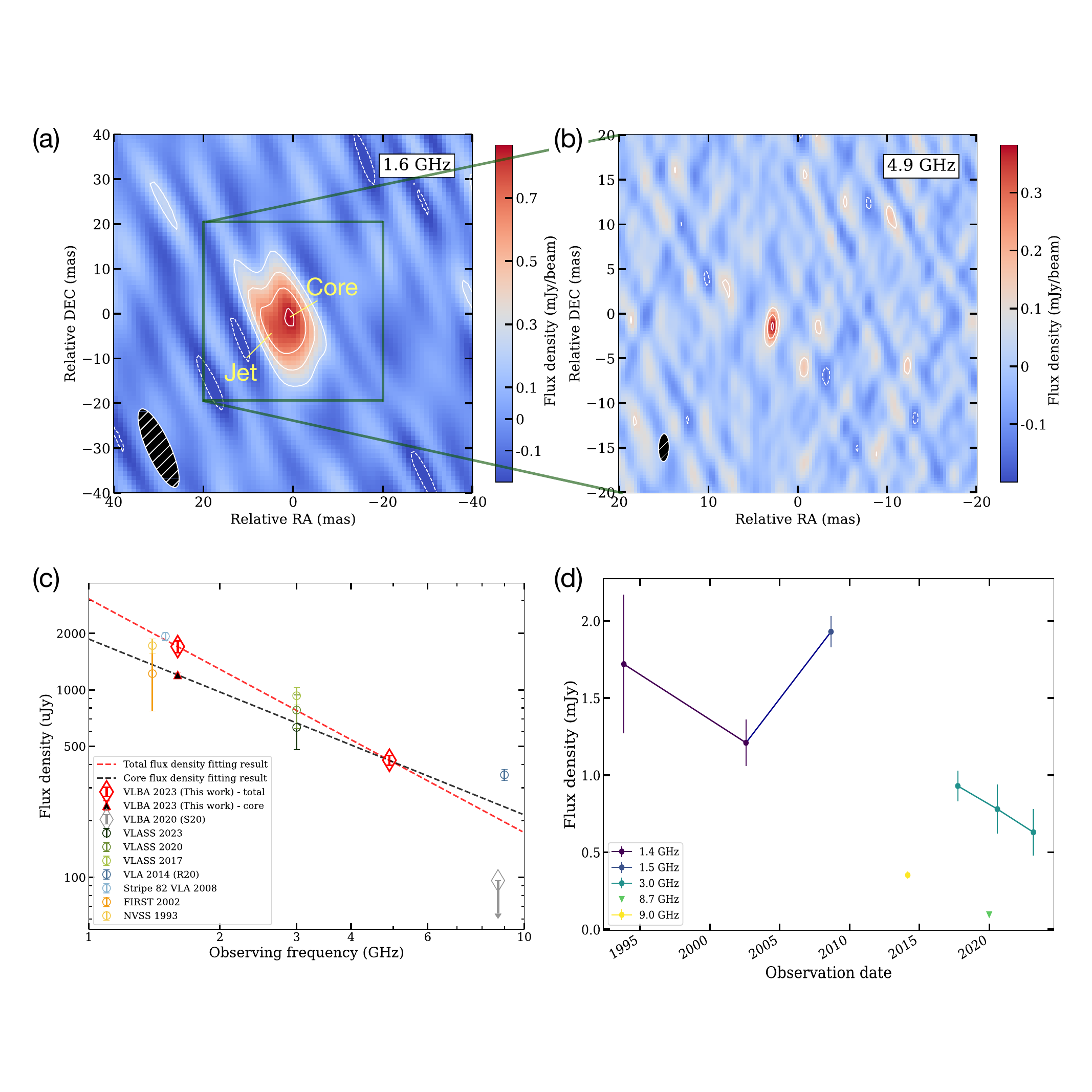}
    \caption{VLBA images in L-band (1.6 GHz, panel (a)) and C-band (4.9 GHz, panel (b)) , radio spectrum in panel (c), and multi-epoch and multi-wavelength flux density comparison in panel d. Uniform weighting is used in creating both images. The contour levels are $ ($-1$, 1, 2, 3, 4) \times 3 \sigma $, in which the RMS noise is $1\sigma =70 \ \mu$Jy~beam$^{-1}$ and $40 \ \mu$Jy~beam$^{-1}$ for L and C band observations, respectively. The jet component is visible in the L-band image as an extension to the southeast of the core. The synthesis beam is shown in the lower left corner. In panel (c), the VLBA measurements from this work are represented by the red diamonds (total flux density) and triangle (core flux density). Other colored labels are from archive or literature VLA and VLBA data, listed in Table S2 (online). }
    \label{fig:J0106-VLBI}
\end{figure}

\begin{figure*}
    \centering 
    \includegraphics[width=0.85\textwidth]{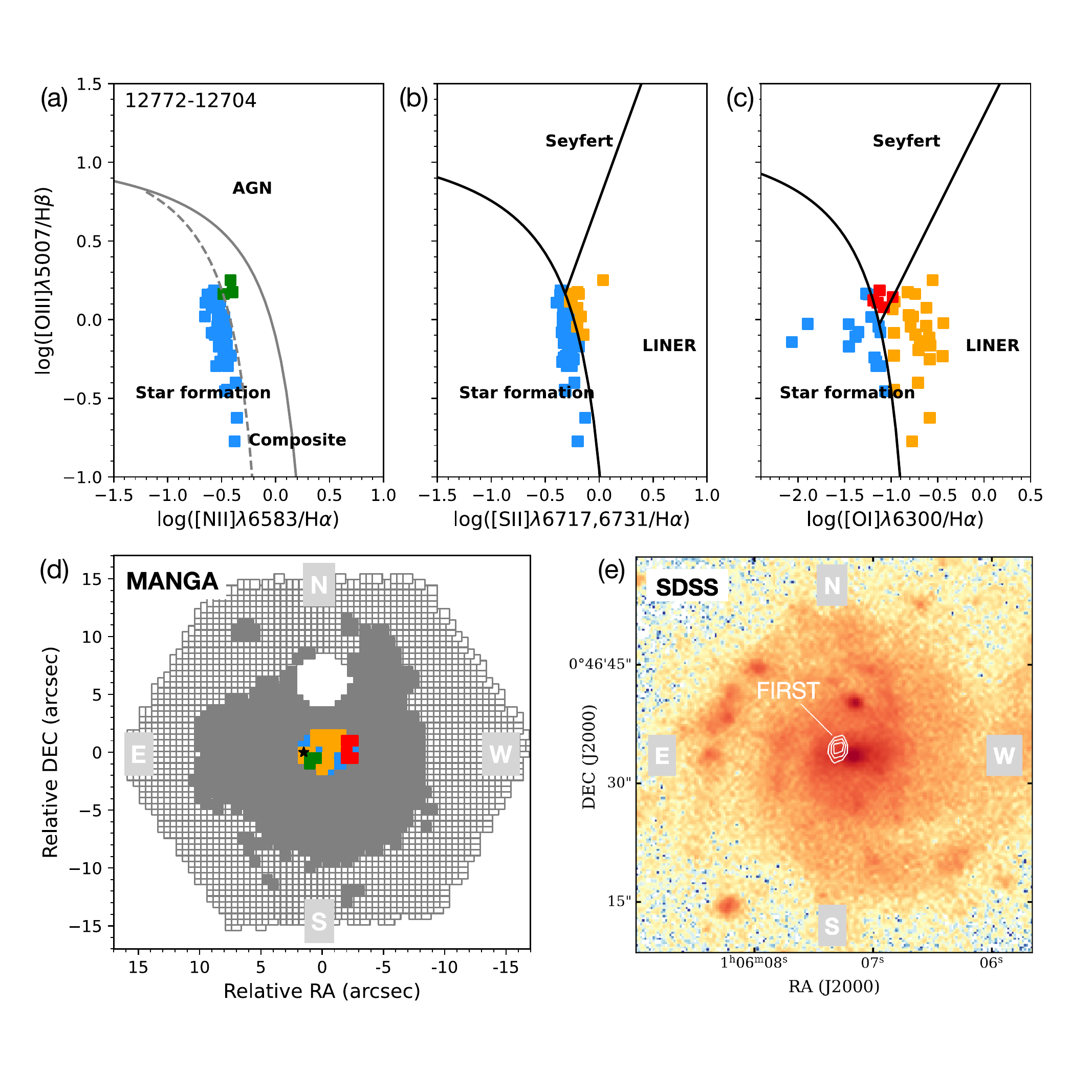}
    \caption{Spatially-Resolved BPT Diagrams of MaNGA 12772-12704 and the optical image. The panels (a),(b), and (c) show BPT diagrams using different emission line ratios, revealing AGN, star formation, and LINER regions. The panel (d) plot shows the spatial distribution of ionization across the galaxy, while the (e) panel presents the SDSS $g$-band image overlaid with radio contours from FIRST survey indicating the off-nuclear source.}
    \label{fig:J0106-BPT}
\end{figure*}

We address these challenges by presenting compelling, multi-faceted evidence for an off-nuclear AGN candidate in the dwarf galaxy MaNGA 12772-12704 using VLBI (Fig. \ref{fig:J0106-VLBI}a and b). The MaNGA integral field unit (IFU) observations \citep{2024MNRAS.528.5252M} reveal that this galaxy exhibits AGN-like line ratios, with emission lines falling within the Seyfert and LINER regions of the $\log([\mathrm{O\,III}]/\mathrm{H}\beta)$ versus $\log([\mathrm{O\,I}]/\mathrm{H}\alpha)$ diagnostic diagram (Fig. \ref{fig:J0106-BPT}a--d). Shown in the optical imaging Fig. \ref{fig:J0106-BPT}e, the host galaxy exhibits no visible signs of merger activity, such as tidal tails, double nuclei, or external disturbances, supporting the interpretation of the radio source as a wandering black hole rather than a secondary AGN in a merging system. Additionally, the MaNGA spectral analysis shows the weak optical AGN emission-line region covering the radio core area, while the galaxy centre spaxel does not exhibit separated high-excitation line ratios. This finding further reinforced the off-nuclear AGN scenario over the dual-AGN interpretation.
The absence of He II $\lambda$4686 emission in the AGN-dominated spaxels is unsurprising, as this line is typically weak and detected in only a small fraction of AGNs, particularly in low-luminosity systems such as the one studied here \citep{2023MNRAS.521.1264T}.
Adopting a stellar mass\footnote{the stellar mass value is derived from spectral energy distribution fitting of SDSS and GALEX photometry and listed in the NASA-Sloan Atlas (NSA catalog) version v1\_0\_1:   https://www.sdss.org/dr17/manga/manga-target-selection/nsa/} of $ M_{\mathrm{stellar}} = 1.52 \times 10^9\,M_\odot$, we estimate the black-hole mass 
$\mathrm \log_{10}(M_{\mathrm{BH}}/M_\odot) = 5.54 \pm \text{0.45}$ utilizing the $ M_{\mathrm{BH}} - M_{\mathrm{stellar}}$ scaling relation \cite{2015ApJ...813...82R}. 
The quoted uncertainty accounts for errors in stellar mass, the scaling relation parameters, and the relation's intrinsic scatter.

We carried out deep, high-resolution radio observations of MaNGA 12772-12704 using the VLBA at L band (1.6 GHz) and C band (4.9 GHz) on 13 and 14 February 2023, respectively. The observations were designed to astrometric precision of approximately 0.1 mas and resolve the radio emission structure. We accumulated $\sim 1$~hour of integration time at each band, employing rapid phase-referencing cycles between the target source and a nearby calibrator to mitigate phase errors in the visibility data. A detailed astrometric analysis presented in the \textit{Supplementary Material} confirms a separation of $2.68\arcsec \pm \text{0.47\arcsec}$ between the VLBI radio core (peak feature) and the optical galaxy centre, corresponding to a projected physical offset of $0.94 \pm 0.16$\,kpc.

Our VLBA observations (Table S1 (online)) yield compelling evidence for a jetted AGN: the 4.9 GHz images reveal a compact radio core with a peak flux density of $0.4 \pm 0.04 \,\mathrm{mJy\, beam^{-1}}$ and a brightness temperature  $T_{\rm b} > 1.8 \times 10^9 \,\mathrm{K}$, exceeding the values can be attributed to star formation by orders of magnitude. At 1.6 GHz, a resolved structure extends approximately $6.2 \pm 0.4 \,\mathrm{mas}$ ($\sim$2.2 pc) southeast (PA $\sim 134^\circ$), interpreted as a compact jet due to its alignment with the core and elongation morphology. The overall radio spectrum is steep ($\alpha \approx -1.2$, where $S_\nu \propto \nu^\alpha$), with the jet showing $\alpha_{\rm jet} \approx -2.0$ (optically thin synchrotron emission)  and  the core showing  $\alpha_{\rm core} \approx -1.5$, typical of low-luminosity AGNs. 

Archival Very Large Array (VLA) data (1993--2023, at frequencies of 1.4–1.5 GHz and 3 GHz) exhibit significant flux density variability (Fig. \ref{fig:J0106-VLBI}c and Table S2 (online)). The measured flux densities show variations: the 1.5 GHz flux density rose from $1.21 \pm 0.15$ mJy (2002) to $1.93 \pm 0.10$ mJy (2008), $\sim 4.6\sigma$ significance; then later 3-GHz measurements declined from $0.93 \pm 0.10$ mJy (2017) to $0.63 \pm 0.15$ mJy (2023). The prolonged flux density variability observed over a 30-year period is another key indicator of sustained accretion activity. 
This non-monotonic, decadal variability indicates sustained accretion, incompatible with SNR monotonic decay ($\sim$years timescale) \citep{2002ARA&A..40..387W}.  Such patterns firmly support an AGN origin over transients.

The radio source in MaNGA 12772-12704 exhibits hallmarks of a jetted AGN: a compact core ($T_{\rm b} > 1.8 \times 10^9$ K), a parsec-scale jet, steep-spectrum emission, and decadal variability.
Confined to $2.68^{\prime \prime}\pm 0.47^{\prime \prime}$ (corresponding to $0.94 \pm 0.16 \,\mathrm{kpc}$) at the source's redshift, it is offset from the galactic nucleus, marking the lowest-redshift ($z=0.017$) VLBI-confirmed wandering AGN in a dwarf galaxy to date.
Being the lowest-$z$ source in the sample of Ref. \citep{2022ApJ...933..160S} (ID6, J0106+0046), it escaped detection by the 9 GHz VLBA due to its steep spectrum and the sensitivity limits. This is unlike their four higher-$z$ ($z > 0.02$) background AGN interpretations ($>2$ kpc offsets), our source favors intrinsic association via:
(1) smaller offset ($<1$ kpc), 
(2) spatial coincidence between the radio core and the optical AGN emission-line region, 
and (3) sustained radio variability inconsistent with a background source (contamination probability $\ll 1\%$). This evidence collectively rule out superposition, supporting a displaced accreting MBH. 

The identification of an actively accreting, off-nuclear MBH candidate in the dwarf galaxy MaNGA 12772-12704 challenges conventional black-hole growth models in low-mass galaxies, providing robust observational support for alternative accretion pathways. This has profound implications for early-Universe SMBH formation, AGN feedback, and black-hole demographics.

The $\approx$0.94 kpc offset of this accreting MBH challenges conventional black‑hole growth models. Theory predicts that such wandering MBHs may arise from gravitational‑wave recoil or asymmetric gas accretion \citep{2021MNRAS.505.5129B}. Our VLBI detection of MaNGA 12772–12704 provides direct evidence that an intermediate‑mass black hole can be spatially displaced from its host’s centre.

A recent example in a more massive galaxy is the offset tidal‑disruption event AT2024tvd, where a $\approx 10^6 \ M_{\odot}$ black hole lies $\approx$0.8 kpc from the galaxy’s $\approx 10^8 \ M_{\odot}$ central SMBH \citep{2025ApJ...985L..48Y}. Multi‑epoch radio observations confirm ongoing accretion in that system. 
Together with the AT2024tvd example, our results suggest that black‑hole growth may proceed away from galactic centres. Instead of relying solely on gas funnelled to the nucleus, wandering black holes can accrete distributed gas, providing an alternative pathway to rapid growth and enriching the diversity of AGN phenomena in dwarf galaxies.

The implications are profound especially for the SMBH formation in the early Universe, where traditional models struggle to explain the rapid growth of SMBHs to $\sim 10^9 M_\odot$ by redshift around 7 \citep{2021NatRP...3..732V}. 
Traditional models assume continuous, efficient accretion fueled by gas funneled to the galactic centre. MaNGA 12772-12704 suggests an alternative pathway: black holes may grow through accretion events distributed throughout their host galaxy. This is particularly relevant in the gas-rich, chaotic galaxies in the early Universe. Instead of relying solely on central reservoirs, black holes could accrete ambient gas or merge with smaller ones. Such a model provides a more plausible pathway for rapid early SMBH growth, challenging the primacy of centralized accretion. 

The presence of a compact ($\sim2.2$ pc) radio jet with a kinetic power of around $10^{41} \, \mathrm{erg\, s}^{-1}$ emanating from the off-nuclear MBH prompts a re-evaluation of AGN feedback in low-mass galaxies. 
These observations suggest that even displaced MBHs can drive mechanical feedback, impacting star formation and gas dynamics within their host galaxies. 
The well-defined jet in this off-nuclear AGN shows that accretion disk-jet systems, commonly seen in powerful AGN within massive galaxies, can form and sustain beyond galactic centre, broadening our insights into AGN physics.

This discovery underscores the need for expanded strategies in searching for IMBHs, raising the question of whether MaNGA 12772-12704 is rare or indicative of a common, undetected population of wandering, accreting black holes. Addressing this requires next-generation facilities like the Square Kilometre Array (SKA) and next-generation Very Large Array (ngVLA), whose superior sensitivity and resolution will enable systematic surveys, probing fainter emissions and resolving finer structures in larger galaxy samples. For instance, SKA-mid could detect IMBH radio emission down to $10^{35}$~erg~s$^{-1}$, while ngVLA's high-frequency capabilities resolve sub-parsec jet structures. These future observations could reshape black-hole demographics, mass distributions, growth mechanisms, early SMBH formation, and BH-dwarf galaxy co-evolution. MaNGA 12772-12704 glimpses the transformative insights ahead.

\section*{Conflict of interest}
The authors declare that they have no conflict of interest.

\section*{Acknowledgments}
This research has been supported by the National SKA Program of China (2022SKA0120102, 2022SKA0130103), National Natural Science Foundation of China (12403023), FAST special Program (NSFC 12041301) and China Postdoctoral Science Foundation (2023M733625, 2024T170968).
Y.Q.L is supported by the Shanghai Post-doctoral Excellence Program and Shanghai Sailing Program (23YF1455700). 
T.A. acknowledges the support from the Xinjiang Tianchi Talent Program.
M.M. acknowledges support from the Spanish Ministry of Science and Innovation through the project PID2021-124243NB-C22. This work was partially supported by the program Unidad de Excelencia María de Maeztu CEX2020-001058-M.
This work used resources of China SKA Regional Centre funded by the Ministry of Science and Technology of the People’s Republic of China and the Chinese Academy of Sciences.
The National Radio Astronomy Observatory (NRAO) is a facility of the National Science Foundation operated under a cooperative agreement by Associated Universities, Inc. 
This paper makes use of the VLBA data from the program BA158.

\section*{Author contributions}
Yuanqi Liu was responsible for data processing and analysis, and wrote the manuscript. 
Tao An conceived the overall research idea, supervised the project, and contributed to manuscript preparation. 
Tao An contributed to the study design and manuscript preparation. 
Yingkang Zhang and Jun Yang provided valuable guidance on VLBA data processing. 
Mar Mezcua collected and analyzed the optical data. 
Ailing Wang and Xiaopeng Cheng contributed to discussions and provided comments on the manuscript.

\clearpage

\appendix

\section*{Supplementary Materials}

\textbf{Cosmology Model}

Throughout this paper, we assume a cosmology with
$H_0 = 70~\mathrm{km\,s^{-1}\,Mpc^{-1}}$, $\Omega_{\mathrm{M}} = 0.27$, and
$\Omega_{\Lambda} = 0.73$, corresponding to a scale of
$0.346~\mathrm{kpc\,arcsec^{-1}}$ at the redshift of MaNGA~12772\textendash12704.

\textbf{VLBA observations}

We conducted high-resolution VLBA observations of 12772-12704 at the L band (central frequency of 1.6 GHz) and C band (central frequency of 4.9 GHz). Nine VLBA stations, including the longest baselines, participated in the observations. The observations, executed between 13-28 February 2023 under project code BA158 (PI: Tao An), employed phase-referencing techniques to mitigate atmospheric phase variations and precisely determine absolute source positions \citep{1995ASPC...82..327B}. Our observing strategy consisted of alternating 5-minute target and 1.5-minute phase calibrator scans at L-band, and 4-minute target and 1.5-minute calibrator scans at C-band. This approach optimizes phase stability while maximizing on-source time. The total on-source integration time for 12772-1204 was 65 and 60 minutes at L and C band, respectively. Data were recorded with dual polarizations and two-bit sampling across four spectral windows per band. Data correlation was performed at the VLBA correlator at the National Radio Astronomy Observatory in Socorro, NM, with a 2-second integration time, balancing temporal resolution with data volume \citep{2011PASP..123..275D}.

\textbf{VLBA data reduction}

We conducted data reduction and analysis using the Astronomical Image Processing System (AIPS), employing a ParselTongue pipeline \footnote{VLBI data processing pipeline \url{https://github.com/SHAO-SKA/vlbi-pipeline}} deployed at the China SKA Regional Centre \citep{2019NatAs...3.1030A,2022SCPMA..6529501A}. Our calibration process encompassed instrumental delay corrections, ionospheric corrections, and fringe fitting. We applied both automatic and manual flagging to mitigate Radio Frequency Interference (RFI) and other data anomalies, crucial for maintaining data integrity. Note that, the antenna MK and SC perform badly, as well as the whole if 2 have been flagged, which stronly influenced the noise level. For imaging and self-calibration, we utilized the Caltech Difmap package. We first created {\sc clean} models of the phase calibrators using the {\sc clean} task, followed by self-calibration for both phase and amplitude. These solutions were then applied to the target source. Our rigorous calibration approach resulted in overall uncertainties of less than 5\%, ensuring high-fidelity images.

To detect radio sources within the VLA observation uncertainty, we mapped all targets within a $\sim 2\arcsec$ radius surrounding the phase center using natural weighting. For the sole detected source, 12772-12704, we shifted the detection to the phase center in the uv-plane before applying the {\sc clean} task to produce the final map. This approach maximizes sensitivity to extended emission while maintaining optimal resolution. Our comprehensive observational and data reduction strategy enables a thorough investigation of the radio properties of our sample, particularly focusing on the intriguing source 12772-12704.

\begin{table}[h]
\caption{VLBA observation parameters and results}
\label{tab:sample}
\begin{tabular}{lll}
\hline
12772-12704 & L band & C band \\
\hline
date & 2023 Feb 13 & 2023 Feb 14 \\
frequency (GHz) & 1.6 & 4.9 \\ 
exposure time (min) & 65 & 60 \\
restoring beam (mas $\times$ mas)  & 19.1 $\times$ 5.2 & 3.1 $\times$ 1.2\\
core RA$^a$\label{note:a} & 01:06:07.3077 ($\pm$0.4) & 01:06:07.3078 ($\pm$0.05)  \\
core DEC$^a$ &  00:46:34.3202 ($\pm$0.5) & 00:46:34.3185 ($\pm$0.1)\\
peak flux density (mJy beam$^{-1}$) &  1.2 $\pm$ 0.07 & 0.4 $\pm$ 0.04  \\
integrated flux density (mJy) & 1.7 $\pm$ 0.27  &  0.4 $\pm$ 0.04  \\
core size (mas) & $<4.3^b$\label{note:b} & $<0.3^c$\label{note:b} \\
brightness temperature (K) &  $>$ 4.7 $\times 10^7$ & $>$ 1.8 $\times 10^9$ \\
\hline
\end{tabular} \\
Note: \hyperlink{note:a}{a}. error in the unit of mas. 

\hyperlink{note:b}{b}. deconvolved size for the core component. 

\hyperlink{note:c}{c}. Minimum resolvable size for the point source. 

\end{table}


\textbf{Brightness temperature}

The rest-frame brightness temperature is calculated with the formula \citep{2005ApJ...621..123U}:

\[ T_{\rm b} = 1.22 \times 10^{12} (1+z) (S / \theta^2) \nu^{-2} [\rm K] \]

where $z$ is the redshift of the source, $S$ is the flux density of the component in Jy, $\theta$ is the source angualr size in mas, and $\nu$ is the observing frequency in GHz.
For unresolved core components, we derived conservative lower limits for $T_{\rm b}$ using the upper bounds of source sizes following Kovalev et al. \citep{2005AJ....130.2473K}. The core sizes were determined to be $<4.3$ mas at 1.6 GHz and $<0.3$ mas at 4.9 GHz, corresponding to brightness temperature limits of d $T_{\rm b} > 4.7 \times 10^7$ K and $\rm T_{\rm b} > 1.8 \times 10^9$ K respectively.

\textbf{Jet Kinetic Power}

The jet kinetic power ($P_{\rm jet}$) is estimated from the rest-frame 1.4 GHz luminosity ($P_{\rm 1.4GHz}$) using the empirically relation from Cavagnolo et al. \citep{2010ApJ...720.1066C}:
\[
P_{\rm jet} \approx 5.8 \times 10^{43}(P_{\rm 1.4GHz}/10^{40})^{0.70} [\rm \ erg \ s^{-1}] .
\]

The value $\mathrm P_{1.4GHz} \sim 1.9 \times 10^{37} \rm erg \ s^{-1}$ is derived from the dual-frequency VLBA observations and k-correction, yielding $P_{\rm jet} \sim 10^{41} \rm \ erg \ s^{-1}$. This estimate is consistent with updated scaling relations for low-luminosity AGN, which incorporate similar power-law forms but adjust for low-power regimes and multi-wavelength constraints. The uncertainty propagates from both the flux measurement ($\sim10-20\%$ based on VLBA errors) and the scaling relation's intrinsic scatter ($\sim0.5$ dex), resulting in an overall uncertainty of approximately $\pm0.5$ dex on $P_{\rm jet}$.

\textbf{Black Hole Mass}

The supermassive black hole mass is estimated using the $M_{\rm stellar}$--$M_{\rm BH}$ scaling relation from Reines et al. \citep{2015ApJ...813...82R}:
\[
\log\left(\frac{{M_{\mathrm{BH}}}}{{M_{\odot}}}\right) = \alpha + \beta \, \log\left(\frac{M_{\rm stellar}}{10^{11} M_{\odot}}\right) ,
\]
using $\alpha = 7.45 \pm 0.08$, $\quad \beta = 1.05 \pm 0.11$.
Propagating the measurement uncertainties from stellar mass and correlation parameters, the scaling relation parameters ($\alpha, \beta$), and the intrinsic scatter of the relation, we obtain the estimate of $\log_{10} M_{\rm BH} = 5.54 \pm 0.45$, accounting for stellar mass uncertainties ($\pm0.15$ dex), scaling relation parameters ($\pm0.3$ dex), and intrinsic scatter ($\pm0.3$ dex).

\textbf{Positions and Offset Estimate}

The spatial offset between the AGN and the host galaxy center was determined using high-precision astrometry from our VLBA observations and the SDSS.
\begin{itemize}
    \item Radio Position (VLBI Core): The position of the compact radio core was measured from the 4.9 GHz VLBA image, which has the highest resolution. The position is RA = 01:06:07.3078, Dec = +00:46:34.3185 (J2000). The astrometric uncertainty, dominated by thermal noise, is estimated to be approximately $\pm 0.1$ mas \citep{2014ARA&A..52..339R}.
    \item Optical Position (Galaxy Center): 
    The optical center of the host galaxy, MaNGA 12772-12704, is calculated by photometric centroid method using \textit{Photutils} package\footnote{ \textit{Photutils} package\url{https://github.com/SHAO-SKA/vlbi-pipeline}}. The position is RA = 01:06:07.136, Dec = +00:46:33.572 (J2000), with a statistical uncertainty  of $0.46^{\prime \prime}$ from MCMC simulation. 
    \item Separation: The angular separation between the radio core and the optical galaxy center is calculated to be 2.68 arcseconds. The total uncertainty on this separation, propagating the errors from both radio and optical positions, is $\pm 0.47$ arcseconds. At the galaxy's redshift ($z=0.017$), this corresponds to a projected physical offset of $0.94\pm 0.16$ kpc.
\end{itemize}


\textbf{Radio Spectral Index}

The spectral index ($\alpha_{\text{jet}}$) of the jet component was estimated using the flux density measured at 1.6\,GHz and an upper limit at 4.9\,GHz. The jet is clearly resolved at 1.6\,GHz, and its flux density ($S_{\text{jet},1.6\,\mathrm{GHz}}$) was measured by fitting and subtracting the core component. At 4.9\,GHz, the jet is not resolved from the core. We therefore place a $3\sigma$ upper limit on its flux density ($S_{\text{jet},4.9\,\mathrm{GHz}} < 0.12$\,mJy) based on the map's RMS noise of 40\,$\mu$Jy/beam. Using the relation $S_\nu \propto \nu^{\alpha}$, these values yield a steep spectral index limit of $\alpha_{\text{jet}} \lesssim -2.0$, which we report as $\alpha_{\text{jet}} \approx -2.0$ in the main text.

While steeper than canonical flat-spectrum AGN cores, the steep core spectrum ($\alpha_{\text{core}} \approx -1.5$) is consistent with radio emission being a mixture of optically thick and thin jet emission, as observed in other nearby radio-quiet AGN \citep{Wang2023}.

\textbf{Archive radio observations} 

Table S2 lists the radio observations in literature which are used in Fig. 1c.

To assess the significance of flux density variability between two radio observations, we calculate the variability significance ($\sigma$) using the formula:
\[\sigma = \frac{ |\Delta S| }{ \sqrt{ \sigma_1^2 + \sigma_2^2 }} ,
\]
where $\Delta S$ is the difference in flux densities between the two epochs, and $\sigma_1$ and $\sigma_2$ are the respective uncertainties of these measurements. 
At 1.4 GHz, the flux density increased from 1.21 $\pm$ 0.15 mJy (FIRST, 2002) to 1.93 $\pm$ 0.10 mJy (VLA, 2008). After applying a K-correction (assuming a redshift of 0.017 and a radio spectral index of $-0.7$), the adjusted VLA flux density is approximately 2.03 mJy. The resulting variability significance is about 4.6$\sigma$, indicating a significant change over a 6-year period. The variability significances between these epochs range from approximately $0.68\sigma$ to $1.67\sigma$ at 3 GHz, suggesting no significant variability over these shorter timescales. The pronounced variability at 1.4 GHz over a longer timespan, contrasted with the relatively stable 3 GHz measurements over shorter intervals, suggests that the source exhibits credible long-term variability.

\begin{table}[h]
\caption{Archived radio measurement}
\label{tab:archive}
\begin{tabular}{lllll}
\hline
\textbf{12772-12704} & frequency & observation date & flux density & reference \\
 & GHz &  & mJy  & \\
\hline
NVSS &  1.4 & Nov 1993 & 1.72 $\pm$ 0.45 & \citep{1998AJ....115.1693C}\\
FIRST & 1.4 & Aug 2002 & 1.21 $\pm$ 0.15 & \citep{1995ApJ...450..559B}\\
VLA (13B-372) & 1.5 & Sep 2008 & 1.93 $\pm$ 0.10 & \citep{2020ApJ...895...98E}\\
VLA (14A-220)  & 9 & Feb-May 2014& 0.352$\pm$ 0.024 & \citep{2020ApJ...888...36R}\\
VLASS (epoch 1) & 3 & Oct 2017 &  0.93 $\pm$ 0.10 & \citep{2020PASP..132c5001L}\\
VLASS (epoch 2) & 3 & Aug 2020 &  0.78 $\pm$ 0.16 & \citep{2020PASP..132c5001L}\\
VLASS (epoch 3) & 3 & Mar 2023 &  0.63 $\pm$ 0.15 & \citep{2020PASP..132c5001L}\\
VLBA  & 8.7 & Jan 2020 & $< 0.096 \, (3\sigma) $ & \citep{2022ApJ...933..160S}\\
\hline
\end{tabular}
\end{table}


\begin{thebibliography}{99}
\bibitem{2020ARA&A..58..257G}
Greene JE, Strader J, Ho LC. Intermediate-mass black holes. Annu Rev Astron Astrophys 2020;58:257–312.

\bibitem{2020ARA&A..58...27I}
Inayoshi K, Visbal E, Haiman Z. The assembly of the first massive black holes. Annu Rev Astron Astrophys 2020;58:27–97.

\bibitem{2015ApJ...806..219C}
Comerford JM, Pooley D, Barrows RS, et al. Merger-driven fueling of active galactic nuclei: six dual and offset AGNs discovered with Chandra and Hubble space telescope observations. Astrophys J 2015;806:219.

\bibitem{2016ApJ...829...37B}
Barrows RS, Comerford JM, Greene JE, et al. Spatially offset active galactic nuclei. I. selection and spectroscopic properties. Astrophys J 2016;829:37.

\bibitem{2024MNRAS.528.5252M}
Mezcua M, {Dom{\'\i}nguez S{\'a}nchez} H. MaNGA AGN dwarf galaxies (MAD) - I. A new sample of AGNs in dwarf galaxies with spatially-resolved spectroscopy. Mon Not R Astron Soc 2024;528:5252–5268.

\bibitem{2025arXiv250617769P}
Popkov AV, Kovalev YY, Plavin AV, et al. Dim cores of radio-bright AGN jets: VLBI and Gaia astrometry pinpoint different parsec-scale features. arXiv e-prints 2025.

\bibitem{2004ApJ...607L...9M}
Merritt D, {Milosavljevi{\'c}} M, Favata M, et al. Consequences of gravitational radiation recoil. Astrophys J Lett 2004;607:L9–L12.

\bibitem{2021MNRAS.505.5129B}
Bellovary JM, Hayoune S, Chafla K, et al. The origins of off-centre massive black holes in dwarf galaxies. Mon Not R Astron Soc 2021;505:5129–5141.

\bibitem{2020ApJ...888...36R}
Reines AE, Condon JJ, Darling J, et al. A new sample of (wandering) massive black holes in dwarf galaxies from high-resolution radio observations, Astrophys J 2020;888:36.

\bibitem{2025MNRAS.536..295M}
Mezcua M, {Dom{\'\i}nguez S{\'a}nchez} H. Correction to:MaNGA AGN dwarf galaxies (MAD) - I. A new sample of AGNs in dwarf galaxies with spatially-resolved spectroscopy. Mon Not R Astron Soc 2025;536:295–297.

\bibitem{2020ApJ...898L..30M}
Mezcua M, {Dom{\'\i}nguez S{\'a}nchez} H. Hidden AGNs in dwarf galaxies revealed by MaNGA: light echoes, off-nuclear wanderers, and a new broad-line AGN, Astrophys J Lett 2020;898:L30.

\bibitem{2024ApJ...974...14S}
Sacchi A, {Bogd{\'a}n} {\'A}, Chadayammuri U,et al. X-ray bright active galactic nuclei in local dwarf galaxies: insights from eROSITA, Astrophys J 2024;974:14.

\bibitem{2023MNRAS.521.1264T}
Tozzi G, Maiolino R, Cresci G, et al. Unveiling hidden active nuclei in MaNGA star-forming galaxies with He II $\rm \lambda$4686 line emission. Mon Not R Astron Soc 2023;521:1264–1276.

\bibitem{2022ApJ...933..160S}
Sargent AJ, Johnson MC, Reines AE, et al. Wandering black hole candidates in dwarf galaxies at VLBI resolution. Astrophys J 2022;933:160.

\bibitem{2011Natur.470...66R}
Reines AE, Sivakoff GR, Johnson KE, et al. An actively accreting massive black hole in the dwarf starburst galaxy Henize2-10. Nature 2011;470:66–68.

\bibitem{2024A&A...682A..36H}
Hoyer N, Arcodia R, Bonoli S,et al. Massive black holes in nuclear star clusters. investigation with SRG/eROSITA X-raydata. Astron Astrophys 2024;682:A36.

\bibitem{2025ApJ...982...10P}
Pucha R, Juneau S, Dey A, et al. Tripling the census of dwarf AGN candidates using DESI early data, Astrophys J 2025;982:10.

\bibitem{2015ApJ...813...82R}
Reines AE, Volonteri M, Relations between central black hole mass and total galaxy stellar mass in the local universe, Astrophys J 2015;813:82.

\bibitem{2002ARA&A..40..387W}
Weiler KW, Panagia N, Montes MJ, et al. Radio emission from supernovae and Gamma-ray bursters. Annu Rev Astron Astrophys 2002;40:387–438.

\bibitem{2025ApJ...985L..48Y}
Yao YH, Chornock R, Ward C, et al. A massive black hole 0.8 kpc from the host nucleus revealed by the offset tidal disruption event AT2024tvd, Astrophys J Lett 2025;985:L48.

\bibitem{2021NatRP...3..732V}
Volonteri M, Habouzit M, Colpi M, The origins of massive black holes. Nat Rev Phys 2021;3:732–743.



\end{thebibliography}

\begin{thebibliography}{99}

\bibitem{1995ASPC...82..327B}
Beasley AJ, Conway JE. VLBI Phase-Referencing. Astron Soc Pac Conf 1995, p. 327.
 
\bibitem{2011PASP..123..275D}
Deller AT, Brisken WF, Phillips CJ, et al. DiFX-2: a more flexible, efficient, robust, and powerful software correlator. Publ Astron Soc Jpn 2011;123:275.

\bibitem{2019NatAs...3.1030A}
An T, Wu XP, Hong XY,SKA data take centre stage in China. Nature Astron 2019;3:1030–1030.

\bibitem{2022SCPMA..6529501A}
An T, Wu XC, Lao BQ, et al. Status and progress of China SKA Regional Centre proto type. Sci China Phys Mech Astron 2022;65:129501.

\bibitem{2005ApJ...621..123U}
Ulvestad JS, Antonucci RRJ, Barvainis R. VLBA imaging of central engines in radio-quiet quasars. Astrophys J 2005;621:123–129.

\bibitem{2005AJ....130.2473K}
Kovalev YY, Kellermann KI, Lister ML. Sub-Milliarcsecond Imaging of Quasars and Active Galactic Nuclei. IV. Fine-Scale Structure. Astron J 2005;130:2473-2505.


\bibitem{2010ApJ...720.1066C}
Cavagnolo KW, McNamara B R, Nulsen PEJ, et al. A relationship between AGN jet power and radio power. Astrophys J 2010;720:1066–1072.

\bibitem{2015ApJ...813...82R}
Reines AE, Volonteri M, Relations between central black hole mass and total galaxy stellar mass in the local universe. Astrophys J 2015;813:82.

\bibitem{2014ARA&A..52..339R}
Reid MJ, Honma M, Microarcsecond radio astrometry, Annu Rev Astron Astrophys 2014;52:339–372.

\bibitem[]{Wang2023}
Wang AL, An T, Zhang YK, et al. VLBI Observations of a sample of Palomar-Green quasars II: characterizing the parsec-scale radio emission. Mon Not R Astron Soc 2023; 525: 6064–6083. 

\bibitem{1998AJ....115.1693C}
Condon JJ, Cotton WD, Greisen EW, et al. The NRAO VLA Sky Survey. Astron J 1998;115:1693–1716.

\bibitem{1995ApJ...450..559B}
Becker RH, White RL, Helfand DJ, The FIRST Survey: Faint Images of the Radio Sky at Twenty Centimeters. Astrophys J 1995;450:559.

\bibitem{2020ApJ...895...98E}
Eftekhari T, Berger E, Margalit B, et al. Wandering massive black holes or analogs of the first repeating fast radio burst? Astrophys J 2020;895:98.

\bibitem{2020ApJ...888...36R}
Reines AE, Condon JJ, Darling J, et al. A new sample of (wandering) massive black holes in dwarf galaxies from high-resolution radio observations, Astrophys J 2020;888:36.

\bibitem{2020PASP..132c5001L}
Lacy M, Baum SA, Chandler CJ, et al. The Karl G. Jansky Very Large Array Sky Survey (VLASS). Science Case and Survey Design. Publ Astron Soc Pac 2020;132:035001 .

\bibitem{2022ApJ...933..160S}
Sargent AJ, Johnson MC, Reines AE, et al. Wandering black hole candidates in dwarf galaxies at VLBI resolution. Astrophys J 2022;933:160.



\end{thebibliography}
\end{document}